\documentclass{PoS}

\font\tenrsfs=rsfs10 at 12pt
\font\sevenrsfs=rsfs7
\font\fiversfs=rsfs5
\newfam\rsfsfam
\textfont\rsfsfam=\tenrsfs
\scriptfont\rsfsfam=\sevenrsfs
\scriptscriptfont\rsfsfam=\fiversfs
\def\mathscr#1{{\fam\rsfsfam\relax#1}}

\makeatletter

\title{Determining the WIMP mass from a single direct detection experiment}

\ShortTitle{WIMP mass determintation}

\author{\speaker{Anne M. Green}\\
        School of Physics and Astronomy, University of Nottingham,
        University Park, Nottingham NG7 2RD, UK\\
        E-mail: \email{anne.green@nottingham.ac.uk}}

\abstract{The energy spectrum of nuclear recoils 
in Weakly Interacting Massive Particle (WIMP)
 direct detection experiments depends on
  the underlying WIMP mass (strongly for light WIMPs, weakly
  for heavy WIMPs).  We discuss how the accuracy with which the WIMP mass
  could be determined by a single direct detection experiment depends on the
  detector configuration and the WIMP properties. In particular we  examine the
  effects of varying the underlying WIMP mass, the
  detector target nucleus, exposure, energy threshold and maximum
  energy, the local velocity distribution and the background event rate and
  spectrum.}

\FullConference{Identification of dark matter 2008\\
		 August 18-22, 2008\\
		 Stockholm, Sweden}

\begin{document}

\section{Introduction}

The direct detection of WIMPs in the lab would not only directly
confirm the existence of dark matter but would also allow us to probe
the WIMP properties, in particular its mass. This would shed light on
its nature and probe extensions of the standard model of particle
physics. Furthermore definitive detection of the WIMP may well require
consistent signals (i.e. with the same inferred WIMP properties) from
direct detection, indirect detection and collider experiments. Here we give a
brief overview of recent work~\cite{green} on determining the WIMP
mass from direct detection experiments (see also
Refs.~\cite{ls,massall,ds}).

The differential event rate (number of events per unit energy, time
and detector mass) has a roughly exponential energy
dependence:~\cite{ls}
\begin{equation}
  \frac{{\rm d}R}{{\rm d} E} \approx 
  c_{1} F^2(E) \left( \frac{{\rm d}R}{{\rm d} E} \right)_{0} 
 \exp{\left(-\frac{E}{c_{2} E_{\rm R}}\right)} \,,
\end{equation}
where $c_{1} $ and $c_{2}$ are fitting parameters of order unity (which
depend on the target mass number and energy threshold), $({\rm d} R/{\rm d}
E)_{0}$ is the event rate in the $E \rightarrow 0 \, {\rm keV}$ limit
and $F(E)$ is the form factor. The characteristic energy
scale of the exponential, $E_{\rm R}$, depends on the WIMP mass,
$m_{\chi}$, and is given by
\begin{equation}
E_{\rm R} = \frac{ 2 m_{A} m_{\chi}^2 v_{\rm c}^2}{(m_{\chi} + m_{A})^2} \,,
\end{equation}
where $m_{A}$ is the target nuclei mass and $v_{\rm c}$ is the local
circular speed. For light WIMPs ($m_{\chi} \ll m_{\rm A}$) $E_{\rm R}
\propto m_{\chi}^2$, while for heavy WIMPs ($m_{\chi} \gg m_{\rm A}$)
$E_{\rm R} \sim {\rm const}$. In other words, for light WIMPs the energy
spectrum is strongly dependent on the WIMP mass while for heavy WIMPs
the dependence on the WIMP mass is far weaker. Consequently it should be
easier to measure the mass of light (compared with the target nuclei) WIMPs
than heavy WIMPs.

\section{Monte Carlo simulations}

We have used Monte Carlo simulations to examine how well a SuperCDMS
like detector~\cite{SuperCDMS} could determine the WIMP mass from the
energies of observed WIMP nuclear recoil events. Our benchmark
detector is composed of Ge, has a nuclear recoil energy threshold of
$E_{\rm \, th} = 10 \, {\rm keV}$ and has no upper limit on the recoil
energy. We assume that the detection efficiency is independent of
energy, the energy resolution is perfect, the background is zero, the
form factor has the Helm form and fix the WIMP-proton cross-section to
be $\sigma_{\rm p}= 10^{-8} \, {\rm pb}$, a factor of a few below the
current exclusion limits~\cite{exclude}. We assume the local WIMP
speed distribution is Maxwellian and the local density is $0.3 \, {\rm
  GeV \, cm}^{-3}$ and consider (efficiency weighted) exposures of
${\cal E}=3 \times 10^{3}$, $3 \times 10^{4}$ and $3 \times 10^{5} \,
{\rm kg \, day}$ which correspond, roughly, to a detector with mass
equal to that of the 3 proposed phases of SuperCDMS taking data for a
year with a $\sim 50\%$ detection efficiency. These are, generally,
optimistic assumptions and will therefore give `best case' results. We
discuss the effects of dropping or varying some of these assumptions
in Sec.~\ref{results}, see also Ref.~\cite{green}.

For each WIMP mass, $m_{\chi}^{\rm in}$, and detector configuration we
calculate the probability distribution of the maximum likelihood
estimator of the WIMP mass by simulating $10^{4}$ experiments. We
first calculate the expected number of events, $\lambda$, from the
input energy spectrum. The actual number of events for a given
experiment, $N_{\rm expt}$, is drawn from a Poisson distribution with
mean $\lambda$. We Monte Carlo generate $N_{\rm expt}$ events from the
input energy spectrum, from which the maximum likelihood mass and
cross-section for that experiment are calculated.  Finally we find the
(two-sided) $68\%$ and $95 \%$ confidence limits on the WIMP mass from
the maximum likelihood masses.

\section{Results and discussion}
\label{results}

\begin{figure}
\includegraphics[width=.8\textwidth]{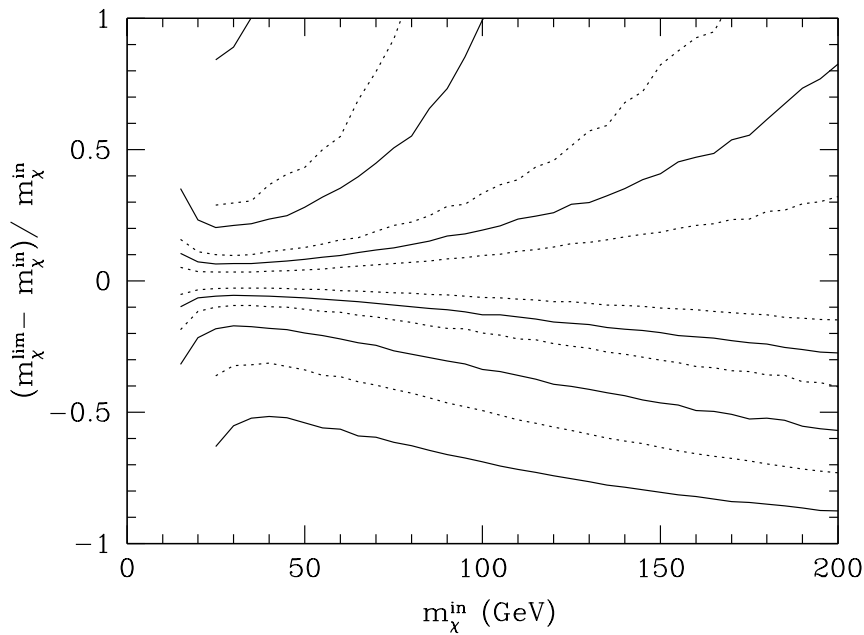}
\caption{The fractional deviation of the WIMP mass limits from the
  input mass, $(m_{\chi}^{\rm lim}-m_{\chi}^{\rm in})/m_{\chi}^{\rm
    in}$, for exposures ${\cal
      E}= 3 \times 10^{3}, 3 \times 10^{4}$ and $3 \times 10^{5} \,
    {\rm kg \, day}$ and input cross-section $\sigma_{\rm p} = 10^{-8}
    \, {\rm pb}$ for the benchmark SuperCDMS like detector. The solid
    (dotted) lines are the 95\% (68\%) confidence limits.  }
\label{fig1}
\end{figure}

The accuracy with which the WIMP mass could be measured by the
benchmark SuperCDMS~\cite{SuperCDMS} like Ge detector described above
is shown in Fig.~\ref{fig1}. With exposures of ${\cal E}= 3 \times
10^{4}$ and $3 \times 10^{5} \, {\rm kg \, day}$ it would be possible
to measure the mass of a light, $m_{\chi} \sim {\cal O}(50 \, {\rm
  GeV})$, WIMP with an accuracy of roughly $25\%$ and $10\%$
respectively. For heavy WIMPs ($m_{\chi} \gg 100 \, {\rm GeV}$) even
with a large exposure it will only be possible to place a lower limit
on the mass.  For very light WIMPs, $m_{\chi} < {\cal O}(20 \, {\rm
  GeV})$, the number of events above the detector energy threshold
would be too small to allow the mass to be measured accurately.

The number of events detected is directly proportional to both the
exposure and the cross-section, therefore these quantities have the
greatest bearing on the accuracy of the WIMP mass determination.

The energy threshold, $E_{\rm th}$, and the maximum energy, $E_{\rm
  max}$, above which recoils are not detected/analysed also affect the
accuracy with which the WIMP mass can be determined. Increasing
$E_{\rm th}$ (or decreasing $E_{\rm max}$) not only reduces the number
of events detected, but also reduces the range of recoil energies and
the accuracy with which the characteristic energy of the energy
spectrum, $E_{\rm R}$, and hence the WIMP mass, can be measured.  For
light WIMPs the small $E_{\rm R}$ means that the expected number of
events decreases rapidly as the energy threshold is increased, while
for heavy WIMPs the large $E_{\rm R}$, and flatter energy spectrum,
means that the smaller range of recoil energies reduces the accuracy
with which $E_{\rm R}$ can be measured. Reducing the maximum energy
only has a significant effect for heavy WIMPs.

The WIMP and target mass dependence of $E_{\rm R}$ suggests that
heavy targets will be able to measure the mass of a heavy WIMP more
accurately, however the rapid decrease of the nuclear form factor with
increasing momentum transfer which occurs for heavy nuclei means that
this is in fact not the case (see also Ref.~\cite{ds}).

If the WIMP distribution on the ultra-local scales probed by direct
detection experiments is smooth, then the $\pm 20 \, {\rm km \,
  s}^{-1}$ uncertainty in the local circular speed~\cite{klb} leads to
a $\sim 10 \%$ systematic error in the determination of $m_{\chi}$.
Changes in the detailed shape of the local velocity distribution lead
to relatively small changes in the shape of the differential event
rate~\cite{drdens}, and hence a relatively small, ${\cal O} (5 \%)$,
systematic uncertainty in the WIMP mass. If the ultra-local WIMP
distribution consists of a finite number of streams, then the energy
spectrum will consists of a number of steps. The positions of the
steps will depend on the (unknown) stream velocities, as well as the
target nuclei and WIMP masses. With multiple targets it would in
principle be possible to constrain the WIMP mass without making any
assumptions about the WIMP velocity distribution~\cite{ds}.

Future experiments aim to have negligible backgrounds, however if the
background rate is not negligible compared with the WIMP event rate it
will be difficult to disentangle a WIMP signal (and the WIMP mass)
from the background if the background spectrum has a similar shape to
the WIMP spectrum (i.e. exponential background, or flat background
and a heavy WIMP). The uncertainties from backgrounds could be
mitigated by using multiple targets and/or using multiple scatter events
to measure/constrain the background spectrum.

\end{document}